\documentclass[twocolumn,aps,prl,superscriptaddress,showpacs,secnumroman,
showkeys]{revtex4}
\usepackage{amsmath,bm}%
\usepackage{graphicx}%

\begin{document}

\title{A Ghoshal-like Test of Equilibration in Near-Fermi-Energy Heavy Ion 
Collisions}

\author{J. Wang}\email[E-mail at:]{wang@comp.tamu.edu}
\author{ T. Keutgen }\thanks{Now at FNRS and IPN, Universit\'e Louvain-Neuve, 
Belgium Catholique de Louvain, B-1348}
\author{ R. Wada }
\author{K. Hagel}
\author{Y. G. Ma}
\thanks{on leave from Shanghai Institute of Nuclear Research,
Chinese Academy of Sciences, Shanghai 201800, China}
\author{M. Murray}
\thanks{Now at University of Kansas, Lawrence, Kansas 66045-7582}
\author{L. Qin}
\author{P. Smith}
\author{J. B. Natowitz}
\affiliation{Cyclotron Institute, Texas A\&M University, College Station, 
Texas 77843}

\author{R. Alfaro}
\affiliation{Instituto de Fisica, Universidad National Autonoma de Mexico, 
Apactado Postal 20-364 01000, Mexico City, Mexico}
\author{J. Cibor}
\affiliation{Institute of Nuclear Physics, ul. Radzikowskiego 152, PL-31-342 
Krakow, Poland}
 \author{A. Botvina}
\affiliation{Cyclotron Institute, Texas A\&M University, College Station, 
Texas 77843}

\author{M. Cinausero}
\affiliation{INFN, Laboratori Nazionali di Legnaro, I-35020 Legnaro, Italy}
\author{Y. El Masri}
\affiliation{FNRS and IPN, Universit\'e Catholique de Louvain, B-1348 
Louvain-Neuve, Belgium}
\author{D. Fabris}
\affiliation{INFN and Dipartimento di Fisica dell' Universit\'a di Padova, 
I-35131 Padova, Italy}
\author{A. Keksis}
\affiliation{Cyclotron Institute, Texas A\&M University, College Station, 
Texas 77843}
\author{S. Kowalski}
\affiliation{Institute of Physics, University of Silesia, PL-40007, Katowice, 
Poland}
\author{ M. Lunardon}
\affiliation{INFN and Dipartimento di Fisica dell' Universit\'a di Padova, 
I-35131 Padova, Italy}
\author{A. Makeev}
\author{N. Marie}
\thanks{Now at LCP Caen, ISMRA, IN2P3-CNRS, F-14050 Caen, France}
\author{ E. Martin}
\affiliation{Cyclotron Institute, Texas A\&M University, College Station, 
Texas 77843}
\author{Z. Majka}
\affiliation{Jagellonian University, M Smoluchowski Institute of Physics, 
PL-30059, Krakow, Poland}
\author{A. Martinez-Davalos}
\author{A. Menchaca-Rocha}
\affiliation{Instituto de Fisica, Universidad National Autonoma de Mexico, 
Apactado Postal 20-364 01000, Mexico City, Mexico}
\author{G. Nebbia}
\author{S. Moretto}
\affiliation{INFN and Dipartimento di Fisica dell' Universit\'a di Padova, 
I-35131 Padova, Italy}
\author{G. Prete}
\author{V. Rizzi}
\affiliation{INFN, Laboratori Nazionali di Legnaro, I-35020 Legnaro, Italy}
\author{A. Ruangma}
\author{D. V. Shetty}
\author{G. Souliotis}
\affiliation{Cyclotron Institute, Texas A\&M University, College Station, 
Texas 77843}
\author{P. Staszel}
\affiliation{Jagellonian University, M Smoluchowski Institute of Physics, 
PL-30059, Krakow, Poland}
\author{M. Veselsky}
\affiliation{Cyclotron Institute, Texas A\&M University, College Station, 
Texas 77843}
\author{G. Viesti}
\affiliation{INFN and Dipartimento di Fisica dell' Universit\'a di Padova, 
I-35131 Padova, Italy}
\author{E. M. Winchester}
\author{S. J. Yennello}
\affiliation{Cyclotron Institute, Texas A\&M University, College Station, 
Texas 77843}
\author{W. Zipper}
\affiliation{Institute of Physics, University of Silesia, PL-40007, Katowice, 
Poland}
\collaboration {\bf The NIMROD collaboration}
\noaffiliation
\author{~and~A.~Ono}
\affiliation{ Department of Physics, Tohoku University, Sendai 980-8578, Japan}

\date{\today}

\begin{abstract}
Calorimetric and coalescence techniques have been employed to probe 
equilibration for hot nuclei produced in heavy ion collisions of 35 to 
55 MeV/u projectiles with medium mass targets. Entrance channel mass 
asymmetries and energies were selected in order that very hot composite 
nuclei of similar mass and excitation would remain after early stage 
pre-equilibrium particle emission. Inter-comparison of the properties 
and de-excitation patterns for these different systems provides evidence 
for the production of hot nuclei with decay patterns relatively independent 
of the specific entrance channel.
\end{abstract}


\maketitle

\section{Introduction }

Determining the extent to which the properties and subsequent de-excitation 
of an excited nucleus depend upon the entrance channel characteristics of 
the collision used to prepare that nucleus is a time honored way of 
exploring equilibration in nuclear systems.  In the early, low energy, 
experiments of Ghoshal~\cite{ghoshal50}, indications that the decay of 
identical compound nuclei, produced at the same excitation energy in 
reactions induced by protons and alpha particles was independent of the 
entrance channel provided very strong evidence for the validity of the 
independence hypothesis~\cite{bohr36,blatt52}, in 
statistical decay theories. In higher energy collisions at near Fermi 
energies, in which faster emission processes remove energy and mass and 
the subsequent composite nucleus undergoes some dynamic evolution prior 
to a possible equilibration, the establishment of equilibration and 
entrance channel independence is more difficult. Nevertheless, it is 
precisely in this domain that nuclei are being investigated to determine 
whether or not their disassembly modes are dynamically or thermodynamically 
driven and, in the latter case, whether they provide evidence of 
liquid-gas phase changes and critical behavior common to other 
liquids~\cite{suraud89,tamain96,chomaz01,elliott03}. Tests 
of the degree of equilibration achieved in these Fermi energy collisions 
are clearly of interest.

In this paper we report on a  Ghoshal-like test of the equilibration of 
medium mass hot composite nuclei produced in central collisions of  
35 MeV/u $^{64}$Zn+ $^{92}$Mo, 40 MeV/u 
$^{40}$Ar + $^{112}$Sn and 55 MeV/u$^{27}$Al + 
$^{124}$Sn  and in mid-central collisions of 47 MeV/u 
$^{64}$Zn + $^{92}$Mo. Correlations between ejectile 
energy and emission time predicted by AMD-V transport model 
calculations~\cite{ono99,wada04,wang04} were employed to follow 
the evolution of the temperatures from the time of first particle emission 
to the time of equilibration. The early temperature evolutions for the four 
different systems show a close correspondence although the time range for 
the observed evolution varies somewhat with reaction system. A very similar 
double isotope ratio temperature, T$_{HHe}\sim6$ MeV, is reached 
at the end of the pre-equilibrium stage of each reaction. The average 
masses and excitation energies remaining at the end of the pre-equilibrium 
stage were determined using calorimetric techniques. The derived masses 
range from 85 to 114 amu and the excitation energies range  from 4.88 to 
5.31 MeV/u. Comparisons of the subsequent evaporation spectra and particle 
multiplicities for the four systems provide strong evidence that the 
de-excitation cascades for these systems are very much alike. The 
combined results indicate that very similar equilibrated hot nuclei are 
produced in the four different collisions. 

\section{Experiment}

The reactions 35 MeV/u $^{64}$Zn+ $^{92}$Mo, 40 MeV/u $^{40}$Ar + $^{112}$Sn, 
55 MeV/u $^{27}$Al + $^{124}$Sn and 47 MeV/u $^{64}$Zn + $^{92}$Mo were 
studied at the K-500 super-conducting cyclotron facility
at Texas A \& M University.  The first three reaction 
systems were selected because various model calculations predict that 
central collisions lead, after pre-equilibrium emission, to hot nuclei 
with very similar composite masses and excitation energies. The 
$^{27}$Al + $^{124}$Sn reaction has a significantly different N/Z ratio 
than the first two reactions. For the fourth reaction, 47 MeV/u 
$^{64}$Zn + $^{92}$Mo, selection of mid-central collisions was predicted 
to lead to a composite system of comparable energy and mass to those of 
the central collisions for the first three systems.  

For these studies we used the NIMROD detector array which consists of a $4\pi$ 
charged particle array inside a $4\pi$ neutron calorimeter. The charged 
particle detector array of NIMROD includes 166 individual CsI detectors 
arranged in 12 rings in polar angles from $\sim4^o$ to $\sim160^o$. In these 
experiments each forward ring included two ``super-telescopes'' composed of 
two Si detectors and 7 Si-CsI telescopes to identify intermediate mass 
fragments (IMF). The NIMROD neutron ball, which surrounds the charged 
particle array, was used to determine the neutron multiplicities by 
thermalization and capture of 
emitted neutrons. As the neutron ball provides little neutron energy or 
angle information, in this experiment we also employed five discrete 
liquid scintillator neutron detectors of the Belgian-French DEMON 
array~\cite{tilquin95} in order to obtain neutron energy and angular 
distributions by time of flight techniques. These neutron detectors 
were placed at laboratory angles of $30^o$ to $110^o$ relative to 
the beam direction and at distances of 2 to 3 meters from the target. 
For these measurements the top quadrant of the mid-section of the neutron 
ball was raised to allow  5 cm gaps between neighboring sections. The DEMON 
detectors viewed the target through one of these gaps. Given the large 
amount of material in the NIMROD detector, this configuration is not 
ideal for measurements of discrete neutron spectra. However the neutron 
ball is a very effective absorber for scattered neutrons. By employing a 
standard reaction, 26 MeV/u $^{40}$Ar + $^{92}$Mo, for which neutron spectra 
and multiplicities have previously been well established by 
Yoshida {\it et al.}~\cite{yoshida92}, we empirically determined the 
required DEMON detector efficiency curves appropriate to our 
experimental configuration. Much greater detail on the detection systems  
and calibrations may be found in references~\cite{wada04,tilquin95}.

During the experiment, data were taken employing two different trigger modes. 
One was a minimum bias trigger in which at least one of the CsI detectors 
detected a particle. The other was a high multiplicity trigger which 
required detected particles in 3 to 5 CsI detectors (depending upon the 
reaction studied). 

\section{Data Analysis}

Many of the techniques applied in this analysis have  been discussed 
previously in greater detail in 
references~\cite{wada04,wang04,cibor00,hagel00,cibor01,ma04}. Only a brief 
summary of these techniques is included in the present work.

An inspection of the two dimensional arrays depicting the detected 
correlation between charged particle multiplicity and neutron multiplicity 
in NIMROD (not shown), reveals a distinct correlation in which increasing 
charged particle multiplicity is associated with increasing neutron 
multiplicity. Although there are significant fluctuations reflecting both 
the competition between the different decay modes and the detection 
efficiencies, these correlations provide a means for selecting the most 
violent collisions. For the analysis reported in this paper we have 
selected events corresponding to the largest observed neutron and charged 
particle multiplicities. For 35 MeV/u $^{64}$Zn+ $^{92}$Mo, 
40 MeV/u $^{40}$Ar + $^{112}$Sn and 55 MeV/u$^{27}$Al + $^{124}$Sn, the 
10\% highest multiplicity events taken with the minimum bias trigger were 
selected in this manner. This selection emphasizes the lower impact 
parameter collisions. For the 47 MeV/u $^{64}$Zn + $^{92}$Mo we selected 
a range of mid-central collisions having charged particle and neutron 
multiplicities similar to those for the events selected for the 
35 MeV/u $^{64}$Zn+ $^{92}$Mo reaction in order to have a comparable 
excitation energy.  

For the selected events we carried out analyses using three source fits to 
the observed energy and angular distributions of the light charged particles. 
The assumed sources are the PLF ( projectile-like fragment )  source, the 
target-like fragment source (TLF) and an intermediate velocity (IV) 
source~\cite{awes81,awes81_2,prindle98,wada89}.  As in the earlier work, 
the IV source typically has a source velocity very close to half of 
that of the projectile, i.e., that expected for nucleon-nucleon scattering. 
From these fits we obtained parameters describing the ejectile spectra 
and multiplicities which can be associated to the three different sources. 
Given the continuous dynamic evolution of the system, such source fits 
should be considered as providing only a schematic picture of the 
emission process.  We have employed them to estimate 
the multiplicities and energy emission at each stage of the reaction.

For the four reactions studied, the parameters for emission of p, d, t,  
$^{3}$He and $^{4}$He from the different  sources,  derived from the fits, 
follow the trends of earlier reported values at such projectile 
energies~\cite{wang04,awes81,awes81_2,prindle98,wada89}. For 
each individual reaction system the IV source slope temperatures for 
p.d,t $^{3}$He and $^{4}$He are quite similar.  They have values in 
the range of 11 to 16 MeV, characteristic of those for pre-equilibrium 
emission in this projectile energy 
range~\cite{wang04,awes81,awes81_2,prindle98,wada89}.  The slope parameters 
for the TLF sources are much lower, in the range of 4-6 MeV. For both sources 
the observed spectra result from a summation of the spectra of particles 
emitted over a range of time. Thus the observed slope temperature values 
are affected by the relative emission probabilities over that time period.

\section{Excitation Energy Determinations}

The masses and excitation energies of the hot nuclei which remain after 
the early (PLF and IV source) emission have been determined from the source 
fit parameters. The masses were obtained by subtracting, from the total 
entrance channel mass, the mass removed by projectile source and 
intermediate source particles. These were determined from integrations 
of the source fits over $4\pi$.  

For each reaction system the excitation energy remaining in the TLF source 
was determined using calorimetric techniques. The data allow two different 
methods to make such a determination. The first is by subtraction, from the 
initial available energy, of  the total energy (kinetic energy and Q value) 
removed by the PLF and IV sources and the kinetic energy of the primary 
remnant. For this purpose, the primary remnant kinetic energy was determined 
from the TLF source velocity and the mass not removed by the IV and 
projectile like sources.  The second method consists of a reconstruction 
of the TLF excitation energy by using the information on the multiplicities 
and kinetic energies of light charged particles, neutrons and intermediate 
mass fragments associated with the TLF source.  
\begin{equation}
E^* = \sum_i\bar{M}_{cp}(i)\bar{E}_{cp}(i)+\bar{M}_n\bar{E}_n + Q + E_\gamma
\end{equation}
where the sum extends over charged particle type i, $\bar{M}_{cp}(i)$ and 
$\bar{M}_n$  are the average  multiplicities of charged particles and 
neutrons,  $\bar{E}_{cp}(i)$ and $\bar{E}_n$ are their average kinetic 
energies, $E_\gamma$ is the total 
energy of the emitted gamma rays (estimated as 10 MeV) and Q is the Q 
value for the observed de-excitation starting from the primary TLF and 
assuming IMF masses consistent with the EPAX parameterization~\cite{Epax}. 
Some differences were observed in the results from the two different methods 
of excitation energy determination. These reflect inefficiencies in IMF 
detection resulting from incomplete coverage with Si detectors, energy 
thresholds and emission at small laboratory angles not subtended by the 
detectors. As the two methods should agree we employed an iterative analysis 
to achieve convergence for the two different techniques, using the missing 
IMF multiplicity as a variable. For this purpose, the average masses and 
kinetic energies of these missing IMF were estimated from results of 
earlier experiments, on some of the same 
systems~\cite{hagel00,wang04,kowalski04}. 

A summary of the mass and excitation energy determinations for the TLF 
sources in these four systems  is presented in Table 1.  The derived 
excitation energies range from 4.88 to 5.31 MeV/u with estimated overall 
uncertainties of $\pm10$\%. In the following discussion we take the 
derived excitation energies to be those of the equilibrated expanded 
systems~\cite{natowitz02,sobotka04}, at the time corresponding to the latest 
particle emission associated with the intermediate source and thus 
appropriate to the temperature determined at that time.

\section{Temperatures at the End of Pre-Equilibrium Emission}

We recently reported on the use of double isotope yield ratio measurements 
to determine the temperature evolution in intermediate energy heavy ion 
collisions~\cite{wang04}. At intermediate energies the observed spectral 
slope parameters derived from the source fits represent only some weighted 
average values as the observed spectra are convolutions of the spectra at 
different emission times. We have therefore employed isotope yields to 
determine the temperatures at the end of the pre-equilibrium emission 
stage of the reaction. Using the same techniques as in 
reference~\cite{wang04} we have determined the double isotope yield ratio 
temperatures as a function of ejectile velocity for the four different 
systems under consideration.The velocity employed is the ``surface 
velocity'', V$_{surf}$, of the emitted particles, defined as the 
velocity of an emitted species at the nuclear surface, prior to 
acceleration in the Coulomb field~\cite{awes81}.  The energy prior 
to Coulomb acceleration is obtained in our analysis by subtraction of 
the Coulomb barrier energy derived from the source fits.  Since the 
early emitted light particle energies are strongly correlated with 
emission times, and the evaporative or secondary emission contributions  
to the spectra are primarily at the lower kinetic energies, the yields 
of higher energy particles are relatively uncontaminated by later 
emission processes.

The double isotope ratio temperatures employed are T$_{HHe}$, derived from 
the yields of d, t, $^3$He and $^4$He clusters.  To calibrate the time-scale 
associated with our data we have used the  results of  AMD --V transport 
model calculations~\cite{wada04} to determine the relationship between 
calculated emission time  and average ejected nucleon velocities  for 
the different systems. The small charged ejectiles - d, t, $^3$He and 
$^4$He, emitted at early times are treated as resulting from coalescence 
of the nucleons ~\cite{awes81,cibor00,hagel00,wang04}. To focus on the earlier 
evolution of the system we selected such clusters emitted at mid-rapidity, 
i.e., at angles of 70 to 80 degrees in the IV source frame.  In this way 
we attempted to isolate the emission associated with the IV source which 
occurs during the thermalization stage of the reaction. At mid-rapidity 
there is little contribution from the PLF source remaining. There is, 
however some observed contribution from the TLF source at low velocities 
in the IV source frame. In the following analysis yields assigned to the 
TLF source have been subtracted from the experimental yields.

As in reference~\cite{wang04}, the AMD-V model 
calculations~\cite{ono99,wada04} indicate a significant slowing in the 
rate of kinetic energy change near a velocity of 3.5 cm/ns. This signals 
the end of the IV (or pre-equilibrium) emission stages and entry into the 
region of slower nuclear de-excitation 
modes, i.e., evaporation, fission and/or fragmentation. At that point the 
sensitivity of the emission energy to time is significantly reduced. 
Therefore we take the temperature at the time corresponding to the 
velocity of 3.5 cm/ns to be that of the hot nucleus at the beginning 
of the final statistical emission stage. 

We present, in Figure 1, the double isotope ratio temperatures as a 
function of emission time.  As time increases, each of the temperature 
evolution curves rises to a maximum and then decreases. Maximum temperatures 
of 10-15 MeV are observed at times in the range of 95 to 110 fm/c. After 
that the temperatures drop monotonically. Since the transport model 
calculations indicate that contributions from the late stage evaporation 
become dominant below V$_{surf} = 3.5$ cm/ns 
(see also reference~\cite{wang04}) we terminate the curves at the times 
corresponding to that velocity.  We take these times to represent the 
end of the IV (or pre-equilibrium) emission stage.  It is worth noting 
that the corresponding temperatures, presented in column 4 of Table 1, 
are near 6 MeV and thus very similar to the limiting temperatures 
previously derived from a systematic investigation of caloric curve 
measurements ~\cite{natowitz02}, in this mass region. 

\begin{figure}
\includegraphics[scale=0.45]{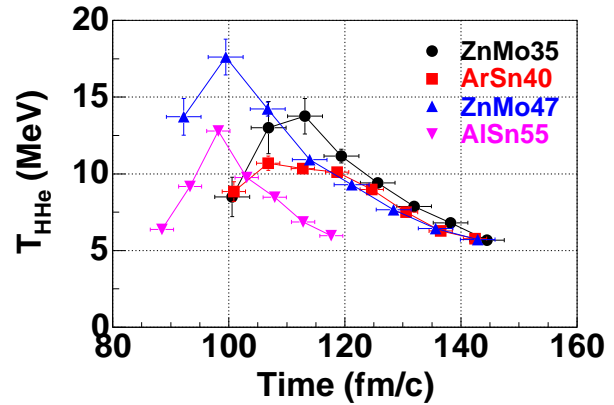}
\caption{Time evolution of the double isotope ratio temperature, T$_{HHe}$,  
for the four different reactions. 35 MeV/u $^{64}$Zn+ $^{92}$Mo- red diamonds,
40 MeV/u $^{40}$Ar + $^{112}$Sn-  orange triangles, 55 MeV/u 
$^{27}$Al + $^{124}$Sn- yellow circles and  47 MeV/u 
$^{64}$Zn + $^{92}$Mo --   blue squares.
}
\label{doubleisotope}
\end{figure}

The trends in Figure 1 are very similar to those reported in 
reference~\cite{wang04}, for the temperature evolution in the reactions of 
26-47 MeV/u $^{64}$Zn projectiles with Ni, Mo and Au targets. The possible 
interpretation of these trends is extensively discussed in that reference. 
It is concluded that the data are consistent with chemical and thermal 
equilibration of the ensemble of events sampled, at least on a local basis. 

If thermal and chemical equilibrium are achieved by the time of entry into 
the final fragmentation or evaporation phase of the reaction and the density 
at that point is not too high~\cite{majka97} an agreement between the thermal 
temperature and the double isotope ratio temperature T$_{HHe}$ at that time 
can be expected. The TLF source fit slope parameters, T$_{slope}$ for alpha 
particles are seen in column 5 of Table 1 to be slightly lower 
than the latest time T$_{HHe}$ temperatures. This is 
not surprising since these T$_{HHe}$ temperatures represent the temperature 
at the start of the de-excitation cascade while the TLF source fits to the 
alpha spectra can be expected to return only an apparent temperature 
reflecting the entire evaporation and cooling history of the TLF source. 
In previous work ~\cite{wada89,hagel88,gonin90} we have found that the 
slope parameters for the alpha particle emission from the TLF source most 
closely approximates the intial thermal temperature of this source, 
reflecting the higher fraction of the alpha emission in the earlier 
part of the de-excitation cascade. The agreement between these thermal 
fit apparent temperatures and the late time chemical temperatures obtained 
from the isotope ratio temperatures, as presented in column 4 is quite 
reasonable given the associated uncertainties.

\begin{table*}[t]
\begin{tabular}{|c|c|c|c|c|c|c|}
\hline
Reaction & A$_{comp}$ & E$^*$ & T$_{HHe}$ & T$_{slope}$ &
$\Delta$A$_{evap}$ & E$^*$/$\Delta$A\\
         &            &MeV/u  & MeV       &MeV          &
                   &MeV/u \\
\hline
35 MeV/u $^{64}$Zn+$^{92}$Mo & $96.4\pm8.7$ & $4.88\pm0.49$ & $5.93\pm0.59$ &
$5.10\pm0.50$ & $38.7\pm3.5$ & $12.1\pm1.7$ \\
35 MeV/u $^{40}$Ar+$^{112}$Sn & 107$\pm9.6$ & $4.98\pm0.50$ & $5.98\pm0.60$ &
$5.80\pm0.30$ & $41.0\pm3.7$ & $13.0\pm1.8$ \\
47 MeV/u $^{64}$Zn+$^{92}$Mo & $85.4\pm7.8$ & $5.14\pm0.51$ & $5.60\pm0.56$ &
$5.50\pm0.03$ & $35.5\pm3.3$ & $12.4\pm1.8$ \\
55 MeV/u $^{27}$Al+$^{124}$Sn & $114\pm10$ & $5.31\pm0.53$ & $6.00\pm0.60$ &
$5.60\pm0.10$ & $44.5\pm4.1$ & $13.6\pm1.9$ \\
\hline
\end{tabular}
\caption{Properties of the Hot Nuclei Produced in mid-central collisions.  
Errors are estimated absolute uncertainties.}
\end{table*}

\section{De-Excitation Cascades}
If, in fact, the hot nuclei have achieved equilibration at very similar 
excitation energies and temperatures, this should be reflected in their 
de-excitation patterns. Figure 2 presents the hot nucleus masses, 
excitation energies and $T_{HHe}$ temperatures for the four different systems 
(in row 1 of the Figure) and a comparison of average slope temperatures, 
multiplicities and kinetic energies for  n, p, d, t, $^{3}$He and $^{4}$He 
emission during the  TLF de-excitation. For each of the quantities presented 
in the figure, a dashed line representing the average values obtained  for 
the 35 MeV/u $^{64}$Zn+ $^{92}$Mo, 40 MeV/u $^{40}$Ar + $^{112}$Sn and 47 
MeV/u $^{64}$Zn + $^{92}$Mo reactions ( which lead to nuclei with similar 
N/Z ratios) is also shown. For each individual system, deviations from 
those averages are seen to be relatively small except for the enhanced 
neutron multiplicity   for the 55 MeV/u $^{27}$Al + $^{124}$Sn reaction, 
which is to be expected given the larger total excitation energy and 
significantly higher N/Z of the hot nucleus produced with this neutron 
rich target~\cite{yoshida92}. Since the total excitation energies do 
vary, we present, in column 6 of Table 1, the total mass removed by TLF 
ejectiles and in column 7 , the ratio of total excitation energy to total 
mass removed by the TLF ejectiles. This latter number is seen to be quite 
stable within experimental uncertainties. The average value, 
$12.8\pm 1.8$ MeV, is close to values obtained in an early study of the 
energetics and de-excitation in reactions of 27 MeV/u  $^{40}$Ar with Ag 
and Ho leading to somewhat heavier hot nuclei with similar excitation 
energies~\cite{rivet86}. In that study, which detected heavy residues, 
equilibration and isotropy of emission was inferred from the residue 
velocity and angular distributions. In the present work, isotropy of 
emission from the TLF source is indicated by the success of that 
assumption in the source fitting process. Except for the expected 
additional favoring of neutron emission over charged particle emission for 
the reaction with the $^{124}$Sn target, the very similar de-excitation 
patterns and energetics for the four hot nuclei which were produced in 
somewhat different dynamical conditions, is most easily understood as a 
reflection of statistical emission from very similar equilibrated hot nuclei. 

It is interesting therefore to ask how the values in Figure 2 compare with 
those calculated using current statistical model codes. For this purpose we 
have carried out calculations using both the GEMINI code~\cite{charity88} and 
the Statistical Multi-fragmentation Model code (SMM)~\cite{bondorf95}. Except 
for an exploration of the effect of varying the level density parameter, a, 
these codes were used with normal default parameters and no attempt has been 
made to tailor them to the current data. For these highly excited systems 
the change in the multiplicities and average energies which results from 
varying the GEMINI code level density parameter was found to be rather 
small. The GEMINI results in Figure 2 were obtained with a level density 
parameter, a = A/9. Yoshida et al. found this to be a reasonable average 
value in their study of neutron emission in the reactions 
of $^{40}$Ar + $^{92}$Mo ~\cite{yoshida92}. 

\begin{figure*}[t]
\includegraphics[scale=0.7]{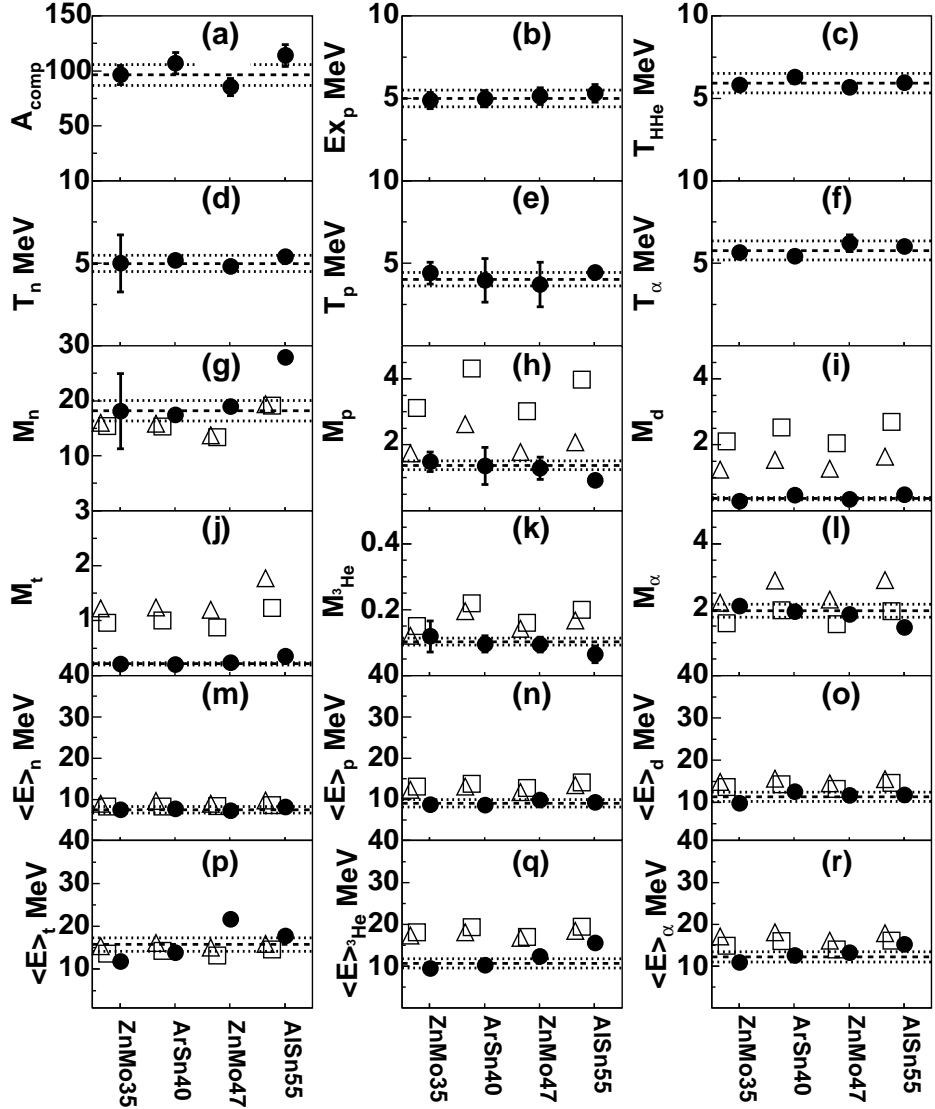}
\caption{Comparison of initial properties of the hot nuclei and de-excitation 
modes for the four different systems.  (a) Masses of the de-exciting hot 
nuclei, (b) Excitation Energies of the hot nuclei, (c) Double Isotope 
Temperatures, T$_{HHe}$, (d) -- (f ) 
Slope temperatures for n,p and , (g) -- (l) multiplicities of n, p, d, t, 
$^3$He and , (m) -- (r) average kinetic energies  of n, p, d, t, $^3$He 
and in the hot nucleus frame. Solid points represent the data for the four 
different systems. Open squares show the results of Gemini 
calculations~\cite{charity88}. Open triangles show the results of SMM 
calculations~\cite{bondorf95}.}
\label{initialproperties}
\end{figure*}

As is seen in that figure, the calculated relative values of the 
multiplicities and average energies for each system show trends quite 
similar to those of the experiments, suggesting that the statistical 
picture has some validity. However, the statistical model calculations 
systematically produce more Z=1 ejectiles and fewer neutrons than have 
been observed in the experiments.  A similar under-prediction of neutron 
emission is seen in reference~\cite{yoshida92}. In addition, except for the 
protons, the calculated average kinetic energies of the charged ejectiles 
are somewhat higher than the experimental values. Given that the dynamic 
calculations indicate that the initial hot de-exciting nuclei produced in 
these collisions have more complicated shapes and density distributions 
than those assumed in the calculations it is perhaps not so surprising 
that some differences between the experimental results and the calculated 
results are observed. 

\section{Conclusion}

Establishing that equilibration has occurred in Fermi-energy collisions is 
difficult,  but remains  an important pre-requisite in determining the 
applicability of many of the techniques which have been utilized or 
proposed as tools to derive information on the nuclear equation of state 
at non normal densities~\cite{suraud89,tamain96,chomaz01,elliott03}.  In this 
paper we have presented results of a Ghoshal-like experimental test of 
equilibration of medium mass hot composite nuclei produced in central 
collisions of 35 to 55 MeV/u projectiles with medium mass targets. 
Although, in contrast to the low energy experiments of Ghoshal, the 
present entrance channel mass asymmetries and energies differ significantly 
for the four different cases studied and the masses of the final residues 
are less than half of the entrance channel masses, comparisons of the 
early temperature evolution and of the subsequent de-excitation patterns 
for the four systems indicate that very similar hot nuclei are indeed 
produced and that they decay in a very similar manner. Multiplicities 
and energies derived from statistical model calculations of the 
de-excitation of these hot nuclei show system to system trends similar 
to those of the experimental results but some quantitative differences 
in the competition between neutron and light charged particle emission.  
Overall, the results are most easily understood as a reflection of 
statistical emission from very similar equilibrated hot nuclei. That 
such systems achieve equilibration is in agreement with the conclusions of 
a recent study in which evidence for equilibration was derived from 
observations of the temperature evolution in similar collisions~\cite{wang04}.

\section{Acknowledgements}

This work was supported by the United States Department of Energy under 
Grant \# DE-FG03- 93ER40773 and by The Robert A. Welch Foundation under 
Grant \# A0330.

\end{document}